\documentclass[twocolumn,superscriptaddress,showpacs,amsmath,footinbib,longbibliography]{revtex4-1}
\usepackage{amsmath}
\usepackage{mathtools} 
\usepackage{graphicx}
\usepackage{dcolumn}
\usepackage{bm}
\usepackage{hyperref}
\hypersetup{
	bookmarks=true,
	colorlinks=true,
	urlcolor=blue,
	citecolor=blue,
	linkcolor=blue, 
	pdftex,
	linktocpage=true, 
	linktoc=all,     
	hyperindex=true
}
\usepackage[mathlines]{lineno}
\usepackage{siunitx} 
\usepackage{comment} 
\usepackage{braket} 
\usepackage{xcolor}

\usepackage{newfloat}
\DeclareFloatingEnvironment[name={Supplementary Figure},fileext=lsf,listname={List of Supplementary Figures}]{suppfigure}

\usepackage[normalem]{ulem}
\newcommand\redout{\bgroup\markoverwith
	{\textcolor{red}{\rule[.5ex]{2pt}{0.4pt}}}\ULon}

\begin{document}
\title{Pattern Formation in Quantum Ferrofluids: from Supersolids to Superglasses}%
\author{J. Hertkorn}%
\affiliation{5. Physikalisches Institut and Center for Integrated Quantum Science and Technology, Universit\"at Stuttgart, Pfaffenwaldring 57, 70569 Stuttgart, Germany
}%
\author{J.-N. Schmidt}%
\affiliation{5. Physikalisches Institut and Center for Integrated Quantum Science and Technology, Universit\"at Stuttgart, Pfaffenwaldring 57, 70569 Stuttgart, Germany
}%
\author{M. Guo}%
\affiliation{5. Physikalisches Institut and Center for Integrated Quantum Science and Technology, Universit\"at Stuttgart, Pfaffenwaldring 57, 70569 Stuttgart, Germany
}%
\author{F. B\"ottcher}%
\affiliation{5. Physikalisches Institut and Center for Integrated Quantum Science and Technology, Universit\"at Stuttgart, Pfaffenwaldring 57, 70569 Stuttgart, Germany
}%
\author{K.S.H. Ng}%
\affiliation{5. Physikalisches Institut and Center for Integrated Quantum Science and Technology, Universit\"at Stuttgart, Pfaffenwaldring 57, 70569 Stuttgart, Germany
}%
\author{S.D. Graham}%
\affiliation{5. Physikalisches Institut and Center for Integrated Quantum Science and Technology, Universit\"at Stuttgart, Pfaffenwaldring 57, 70569 Stuttgart, Germany
}%
\author{P. Uerlings}%
\affiliation{5. Physikalisches Institut and Center for Integrated Quantum Science and Technology, Universit\"at Stuttgart, Pfaffenwaldring 57, 70569 Stuttgart, Germany
}%
\author{T. Langen}%
\affiliation{5. Physikalisches Institut and Center for Integrated Quantum Science and Technology, Universit\"at Stuttgart, Pfaffenwaldring 57, 70569 Stuttgart, Germany
}%
\author{M. Zwierlein}%
\affiliation{MIT-Harvard Center for Ultracold Atoms, Research Laboratory of Electronics, and Department of Physics, Massachusetts Institute of Technology, Cambridge, Massachusetts 02139, USA}
\author{T. Pfau}%
\email{t.pfau@physik.uni-stuttgart.de}
\affiliation{5. Physikalisches Institut and Center for Integrated Quantum Science and Technology, Universit\"at Stuttgart, Pfaffenwaldring 57, 70569 Stuttgart, Germany
}%

\date{\today}

\begin{abstract}
	Pattern formation is a ubiquitous phenomenon observed in nonlinear and out-of-equilibrium systems. In equilibrium, quantum ferrofluids formed from ultracold atoms were recently shown to spontaneously develop coherent density patterns, manifesting a supersolid. We theoretically investigate the phase diagram of such quantum ferrofluids in oblate trap geometries and find an even wider range of exotic states of matter. Two-dimensional supersolid crystals formed from individual ferrofluid quantum droplets dominate the phase diagram at low densities. For higher densities we find honeycomb and labyrinthine states, as well as a pumpkin phase.	We discuss scaling relations which allow us to find these phases for a wide variety of trap geometries, interaction strengths, and atom numbers. Our study illuminates the origin of the various possible patterns of quantum ferrofluids and shows that their occurrence is generic of strongly dipolar interacting systems stabilized by beyond mean-field effects.
\end{abstract}

\maketitle

\section{Introduction}
Classical ferrofluids, which are colloidal suspensions of fine magnetic particles in a fluid, are a model system for self-organized equilibrium \cite{Seul1995,Rosensweig1997,Andelman2009,MorphogenesisBookBourgine2011}. The long-range nature of the magnetic dipolar interaction between their constituent particles allows them to develop macroscopic patterns or textures in equilibrium. These patterns --- also commonly referred to as morphologies --- emerge in a large variety of physical systems irrespective of their microscopic structure and interactions \cite{Seul1995, Andelman2009}. The morphologies notably include droplet (``bubble"), honeycomb (``foam") and labyrinthine (``stripe") phases \cite{Rosensweig1983,Dickstein1993,Seul1995,Jackson1994,Rosensweig1997,Florence1997,Miranda2005,Andelman2009,Zakinyan2017,MorphogenesisBookBourgine2011}. These can be found in equilibrium in systems as diverse as quantum ferrofluids \cite{Lahaye2009,Kora2019,Bottcher2020}, superfluid helium \cite{Grebenev1998,Dalfovo2001,Toennies2001}, the intermediate phase of type-I superconductors \cite{Huebener1974,Cebers2005,Prozorov2007,Prozorov2008}, optically nonlinear media \cite{Ciaramella1993,Edwards1994,Ackemann1995,Arecchi1999,Buckley2004,Labeyrie2014,Maucher2016,Maucher2017,Zhang2018,Zhang2021,Baio2020,Baio2021}, biological matter \cite{Hardenberg2001,Riedel2005,Dunkel2013,Liu2014}, nuclear pasta in ultra-dense neutron stars and white dwarfs \cite{Ravenhall1983,Chamel2008,Caplan2017} as well as in out-of-equilibrium systems \cite{Cross1993} in convection patterns arising from the Rayleigh-Bénard instability \cite{Ahlers1993,Morris1993,Ahlers2009}, and in a plenitude of chemical mixtures displaying reaction-diffusion (``Turing") patterns \cite{Turing1952,Ouyang1991}.

Quantum ferrofluids can be made from strongly dipolar Bose-Einstein condensates (BECs) \cite{Lahaye2007,Lahaye2009, Kadau2016}, which are superfluids in contrast to their classical counterparts \cite{Bismut2012,Wenzel2018}. Atoms in these BECs interact with the same dipolar interaction that has proven to be archetypical of structure formation in equilibrium. The great tunability of interaction strengths in atomic systems \cite{Bloch2008}, the presence of a crystalline droplet phase in classical ferrofluids, and the superfluid nature of quantum ferrofluids have motivated the search for the elusive supersolid phase in dipolar BECs, where crystalline order coexists with global superfluidity \cite{Prokofev2007,Balibar2010,Boninsegni2012}. Consequently, much attention has been given to the droplet morphologies of quantum ferrofluids \cite{Bulgac2002,Santos2003,Jona-Lasinio2013,Petrov2015,Schmitt2016,Kadau2016,Ferrier-Barbut2016,Chomaz2016,Baillie2016,Baillie2017,FerrierBarbut2018GetThin,Baillie2018,FerrierBarbut2019,Bottcher2019droplet,Bottcher2020,Xi2020,Lee2020,Luo2020}. Understanding that these morphologies are stabilized by repulsive quantum fluctuations \cite{Schutzhold2006,Lima2011,Lima2012,Petrov2015,Ferrier-Barbut2016} was crucial for the experimental discovery of elongated dipolar supersolids in cigar-shaped traps \cite{Tanzi2019, Bottcher2019, Chomaz2019, Guo2019, Tanzi2019a}. Despite rapid developments in this field, the dipolar supersolids have been experimentally limited to the droplet morphology and mostly one-dimensional (1D) crystal structures \cite{Tanzi2019, Bottcher2019, Chomaz2019, Guo2019, Tanzi2019a,Bottcher2020, Natale2019, Hertkorn2019, Blakie2020supersolelongate, Hertkorn2021, Roccuzzo2019, Tanzi2021,Blakie2020variational,Pal2020numberfluct,Lee2020,Ilzhoefer2021}, although first steps toward two-dimensional (2D) supersolid droplets have recently been made \cite{Roccuzzo2020,Gallemi2020,Schmidt2021,Norcia2021,Hertkorn2021SSD2D}. In an infinite system, the ground-state phase diagram of 2D arrangements of dipolar supersolids showed honeycomb supersolid structures \cite{Zhang2019}. Earlier studies investigating the potential 2D honeycomb and labyrinthine phases in BECs considered more complex multi-component systems \cite{Saito2009,Kawaguchi2012,Wilson2012,Xi2018} and their dynamical (Rayleigh-Taylor) instabilities \cite{Sasaki2009,Gautam2010,Kadokura2012} or infinite quasi-2D geometries with three-body interactions instead of quantum fluctuations \cite{Lu2015}.

Here, we study single-component quantum ferrofluids confined in cylindrically symmetric geometries, including beyond mean-field quantum fluctuations.  We find that extending the geometry from 1D to 2D in a trapped system extends not only the crystal structure of the droplet phase to the second dimension, but also gives rise to new morphologies. We show that dipolar BECs have a remarkably rich phase diagram as we find quantum liquid states of matter, including supersolid honeycomb and superglass labyrinthine morphologies beyond the supersolid droplet morphology.

In Sec.~\ref{sec:Methods}, we briefly review our methodology and give an overview of the interactions in quantum ferrofluids. We present the ground-state phase diagram of quantum ferrofluids for an oblate trap geometry in Sec.~\ref{sec:Pattern} and discuss the types of morphologies, their location in the phase diagram, and the origin of the pattern formation (morphogenesis). In Sec.~\ref{sec:ScalingProp}, we show that dimensionless units reveal scaling properties of quantum ferrofluids in the presence of quantum fluctuations and discuss the geometry dependence of the patterns. The scaling relations generalize the phase diagram discussed in Sec.~\ref{sec:Pattern} to a wide range of trap geometries and allow to tune the strength of the stabilization mechanism of the morphologies. Furthermore we show that, by simply adjusting the trapping confinement, geometric transitions between BEC, honeycomb, labyrinthine, and droplet states are possible. The characteristic length scale of the patterns follows the same scaling with trapping geometry that is known from the roton momentum of dipolar BECs and extends it to new and unexpected regimes. Finally, we conclude in Sec.~\ref{sec:Conclusion} and provide an outlook of our study.

\section{Methodology}\label{sec:Methods}
A dilute dipolar BEC at zero temperature is described within an effective mean-field theory, provided by the extended Gross-Pitaevskii equation (eGPE) 
\begin{equation}\label{eq:GPE}
	i \hbar \partial_t \psi = \left(\hat{H}_0 + g_s |\psi|^2 + g_\mathrm{dd} (U_\mathrm{dd} * |\psi|^2) + g_\mathrm{qf} |\psi|^3 \right)\psi,
\end{equation}
where the wavefunction $\psi$ is normalized to the atom number ${N=\int \mathrm{d}^3r\, |\psi(\boldsymbol{r},t)|^2}$ and ${\hat{H}_0 = -\hbar^2 \nabla^2 / 2 M + V_\mathrm{ext}(\boldsymbol{r})}$ \cite{Ronen2006,Saito2016,Wenzel2017,Roccuzzo2019}. We consider a cylindrically symmetric harmonic trap ${V_\mathrm{ext}(\boldsymbol{r}) = M \omega_r^2 (x^2 + y^2 + \lambda^2 z^2)/2}$ with aspect ratio ${\lambda = \omega_z / \omega_r}$ and the mass $M$ of the atomic species. The contact and dipolar interaction strengths ${g_s = 4\pi \hbar^2 a_s/M}$ and ${g_\mathrm{dd} = 4\pi \hbar^2 a_\mathrm{dd}/M}$. These quantities are determined by the scattering length $a_s$ and the dipolar length ${a_\mathrm{dd} = \mu_0 \mu_m^2 M / 12 \pi \hbar^2}$ with the magnetic moment $\mu_m$. The long-range and anisotropic dipolar interaction with the dipoles aligned by a magnetic field along the $\hat{\boldsymbol{z}}$-direction is given by ${U_\mathrm{dd}(\boldsymbol{r}) = (3/4\pi)(1-3z^2/r^2)/r^3}$ \cite{Lahaye2009}. The dipolar mean-field potential is given by the convolution ${(U_\mathrm{dd} * |\psi|^2)(\boldsymbol{r},t) = \int\!\mathrm{d}^3r' U_\mathrm{dd}(\boldsymbol{r}-\boldsymbol{r}')|\psi(\boldsymbol{r}',t)|^2}$. Beyond mean-field quantum fluctuations are taken into account within the local density approximation for dipolar systems \cite{Schutzhold2006,Lima2011,Lima2012,Petrov2015,Ferrier-Barbut2016} by the Lee-Huang-Yang (LHY) correction $g_\mathrm{qf}|\psi|^3$ with ${g_\mathrm{qf} \simeq (32/3\sqrt{\pi}) g_s a_s^{3/2} (1+ 3\epsilon_\mathrm{dd}^2/2)}$ and ${\epsilon_\mathrm{dd} = a_\mathrm{dd} / a_s}$ is the relative dipolar strength.

In the following, we are most interested in the ground states of the system for parameters where structured forms of matter arise. To understand structure formation as a result of competing interactions \cite{Seul1995}, we consider the underlying energy contributions of a state described by the eGPE in the context of a density functional theory \cite{Dalfovo2001,Dalfovo1999,Archer2008,Antoine2018,Heinonen2019}. The eGPE can be formulated as ${i \hbar \partial_t \psi = \delta E/\delta \psi^*}$ \cite{PitaevskiiBook2016}, where the right hand side is the functional derivative of the energy functional
\begin{equation}\label{eq:EnergyFunctional}
\begin{split}
E &= \int\! \mathrm{d}^3r  \left( \frac{\hbar^2}{2M}|\nabla\psi|^2 + V_\mathrm{ext} |\psi|^2 \right. \\
&+ \left. \frac{1}{2} g_s |\psi|^4 +\frac{1}{2} g_\mathrm{dd} |\psi|^2 (U_\mathrm{dd} * |\psi|^2) + \frac{2}{5} g_\mathrm{qf} |\psi|^5 \right)
\end{split}
\end{equation}
with respect to $\psi^*$. We find ground states by a direct minimization of Eq.~\eqref{eq:EnergyFunctional} using conjugate gradient techniques \cite{Modugno2003,Ronen2006,Antoine2017,Antoine2018}.

We denote the density $n = |\psi|^2$ and the integrands of Eq.~\eqref{eq:EnergyFunctional} as an energy density $\mathcal{E}$. Equation~\eqref{eq:EnergyFunctional} contains the repulsive contributions by the contact interaction $\mathcal{E}_\mathrm{con} \propto g_s n^2$ and quantum fluctuations $\mathcal{E}_\mathrm{qf} \propto g_\mathrm{qf} n^{5/2}$, which importantly have a distinct scaling with the density \cite{Bulgac2002,Bender2003,Petrov2015,Lu2015,Bottcher2020}. The dipolar interaction is long-range and anisotropic and can give an attractive contribution $\mathcal{E}_\mathrm{dd} < 0$ for particles that arrange in a head-to-tail configuration. The competition between the attractive dipolar and repulsive contact interaction can therefore lead to mean-field instabilities that are stabilized by the stronger density scaling of the quantum fluctuations. Repulsive and attractive interactions at different length and density scales are the key components in Eq.~\eqref{eq:EnergyFunctional} that lead to structure formation and are also present in other systems such as optically coupled cold atoms \cite{Ostermann2016,Zhang2019,Zhang2021,Baio2020,Baio2021}, nuclear matter \cite{Bender2003,Chamel2008}, helium droplets \cite{Dalfovo2001,Toennies2001}, and colloidal systems \cite{Seul1995,Andelman2009,Nelissen2005,Liu2008,Archer2008}. In the context of cold atomic physics, strongly dipolar BECs represent a realistic system holding the potential for complex pattern formation in equilibrium \cite{Lu2015,Kora2019,Zhang2019,Bottcher2020}.

\section{Patterns in quantum ferrofluids}\label{sec:Pattern}
Here we first discuss the various morphologies that can be found in the phase diagram of quantum ferrofluids in oblate traps. Second, we turn to the origin of the pattern formation, the morphogenesis.

We consider a strongly dipolar BEC of $^{162}$Dy atoms (${a_\mathrm{dd} \simeq 130\, a_0}$) confined in a cylindrically symmetric oblate trap with trapping frequencies ${\omega/2\pi = (125,\, 125,\, 250)\,\si{\hertz}}$, aspect ratio $\lambda = 2$, and a magnetic field along $\hat{\boldsymbol{z}}$. The phase diagram for the chosen parameters is connected by scaling relations to similar phase diagrams in other trap geometries or with other atomic species  as we show in Sec.~\ref{sec:ScalingProp}.

\begin{figure}[tb!]
	\includegraphics[trim=0 0 0 0,clip,scale=0.57]{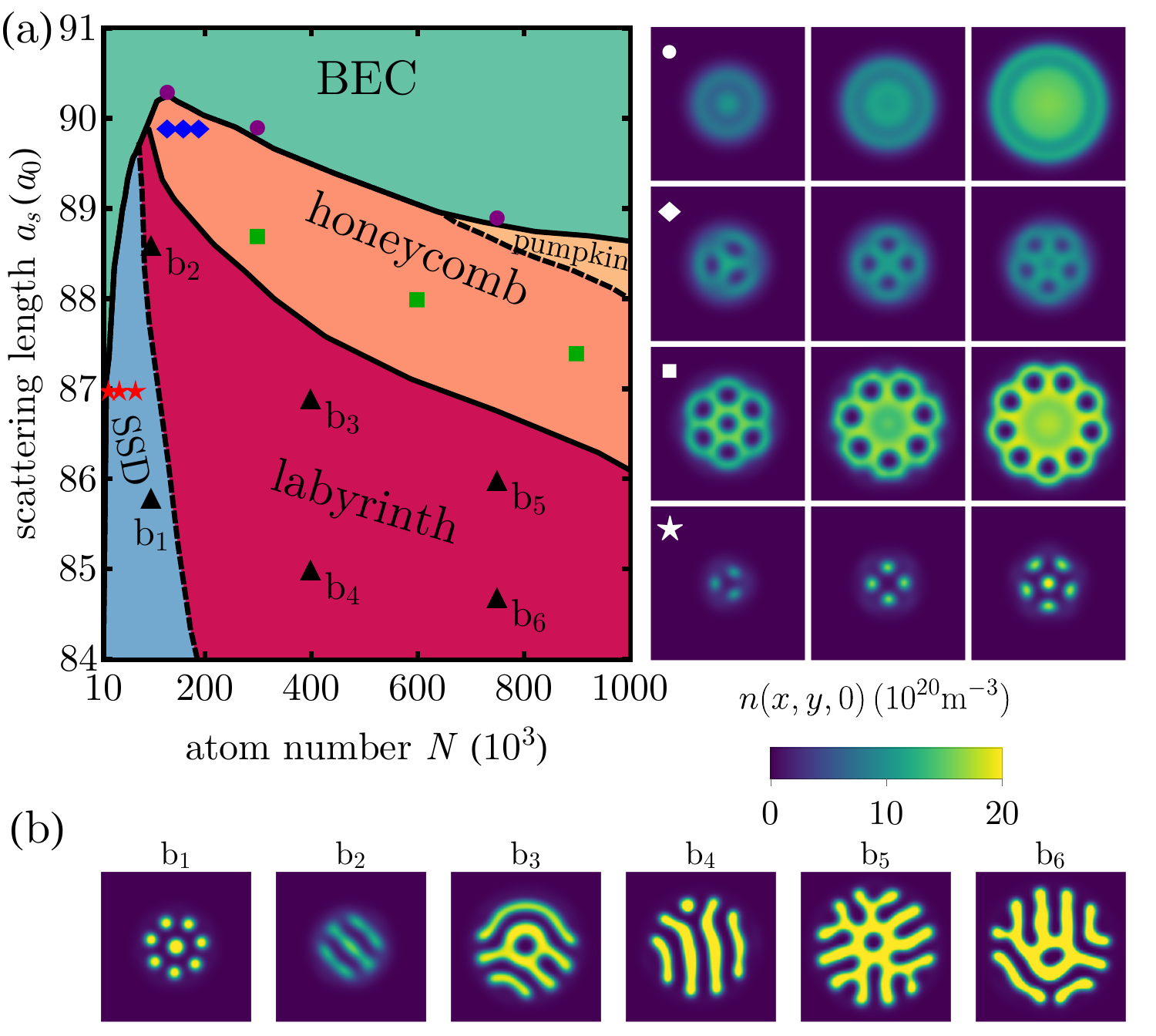}
	\caption{Phase diagram and morphologies beyond the supersolid droplet regime. (a) The left-hand side show the $N$-$a_s$ phase diagram for trap frequencies ${\omega/2\pi = (125,\, 125,\, 250)\,\si{\hertz}}$. The right hand side shows 2D density cuts $n(x,y,0)$ at relevant points in the phase diagram, shown by the corresponding markers. The density distributions for a specific marker are ordered in atom number from left to right. The BEC at high atom numbers has a ring of depleted density near its boundary (circles) and forms honeycomb structures toward smaller scattering lengths (diamonds). The honeycomb structures persist to higher atom numbers (squares) and move outwards to the rim of the density distributions, while the core of the BEC spatially saturates in density. An example of the pumpkin state can be seen in Fig.~\ref{fig:Pumpkin}. Dashed lines indicate crossovers between different regions.
	(b) The transition between droplets and honeycombs occurs via stripes (b$_2$) that break up into droplets at small scattering lengths (b$_1$). At high atom numbers and low scattering lengths (b$_3$-b$_6$), labyrinth structures form that are almost degenerate with many other morphologically different labyrinth structures. The supersolid droplets form density connections toward higher atom numbers and transition to labyrinthine structures. The field of view for the 2D-densities in (a) and (b) is $14 \times 14\,\si{\micro\meter\tothe{2}}$.}
	\label{fig:PhaseDiag}
\end{figure}
In order to gain insight into the pattern formation of quantum ferrofluids we map out the ground-state phase diagram in a wide range of interaction strengths and atom numbers around the instability boundary from a BEC to structured states of matter, as shown in Fig.~\ref{fig:PhaseDiag}. We search for the ground state at every scattering length and atom number by setting a random initial wavefunction \cite{supmat}, allowing us to avoid hysteresis effects when crossing phase boundaries in parameter space \cite{Bottcher2019}.

The boundary below which the BEC transitions to structured phases is described by a critical scattering length $a_{s,c}$. The structured states have a reduced symmetry compared to the rotationally symmetric BEC state, as the continuous rotational symmetry is spontaneously broken for scattering lengths below $a_{s,c}$. The spontaneous rotational symmetry breaking characterizes the appearance of supersolid or superglass phases, where crystalline or amorphous spatial structure coexists with superfluid flow \cite{Boninsegni2012}. We find that the BEC can transition to a variety of patterns, namely supersolid droplet (SSD), honeycomb and stripe or labyrinth phases \cite{Bottcher2020,Kora2019,Zhang2019,Echeverria2020}. The phase diagram is shown in Fig.~\ref{fig:PhaseDiag}(a) on the left hand side and examples of patterns for the different phases are shown on the right hand side and in (b).

As shown in Fig.~\ref{fig:PhaseDiag}(a) (circles), the BEC states near $a_{s,c}$ develop a radial substructure such that they differ from a Thomas-Fermi density distribution. The BEC states in the range $N \simeq 60$-$200 \times 10^3$ near $a_{s,c}$ show a ring of depleted density near their boundary in addition to the maximum density in the center of the trap (Fig.~\ref{fig:PhaseDiag}(a), circles, left column). At intermediate atom numbers ($N \simeq 200$-$400 \times 10^3$) a second minimum in the center of the trap can occur and toward higher atom numbers, the trap center is filled with atoms and only the depleted density ring near the boundary remains (Fig.~\ref{fig:PhaseDiag}(a), circles, right column). A special case of the BEC shape occurs toward lower atom numbers ($N \lesssim 50 \times 10^3$), where the maximum density in the center of the trap vanishes, leaving only the density ring away from the trap center. These states are known as biconcave or blood cell states \cite{Eberlein2005,Ronen2006,Ronen2007,Dutta2007,Wilson2008,Wilson2009,Wilson2009AngularCollapse,Lu2010,Blakie2012,Martin2012,Kawaguchi2012,Bisset2013,Schmidt2021} due to the similarity to the shape of a red blood cell. Indirect experimental evidence of these shapes has recently been found \cite{Schmidt2021} and a theoretical study explained their connection to supersolid droplets by investigating elementary excitations across the transition \cite{Hertkorn2021SSD2D}.

\begin{figure*}[tb!]
	\includegraphics[trim=0 0 0 0,clip,scale=1.025]{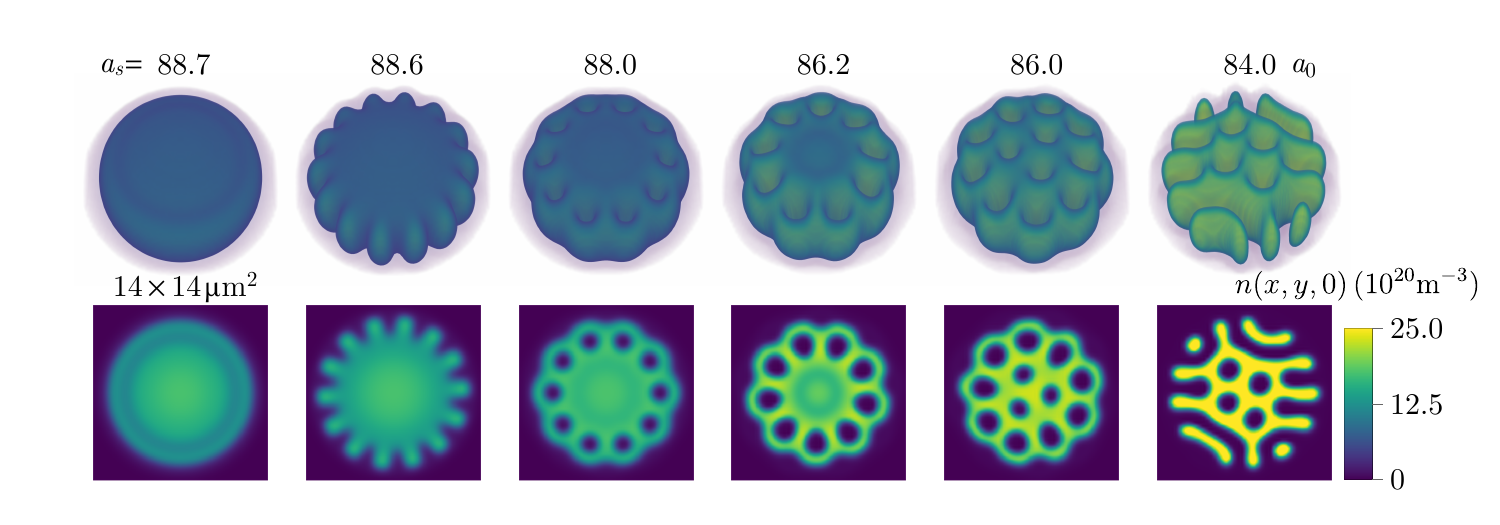}
	\caption{Morphogenesis. Shown are density distributions of the ground state with ${N= 1000 \times 10^3}$ atoms. At such high atom numbers, the BEC features a density saturated core and develops a density depleted ring in its crust when the scattering length $a_s$ is reduced to its first boundary ($a_s = 88.7\,a_0$). For smaller $a_s$, the outer high-density ring becomes unstable and breaks in, yielding the pumpkin-like state ($a_s = 88.6\,a_0$). As $a_s$ is reduced further, the depleted density wanders closer to the core and is closed off by an outer shell of density, yielding a dense core surrounded by a honeycomb structure ($a_s = 88.0\,a_0$). When the depleted density has expanded yet closer to the core a new depleted density ring forms ($a_s = 86.2\,a_0$) which again forms honeycomb structures ($a_s = 86.0\,a_0$). As the stabilizing repulsion becomes insufficient to uphold the fine density bridges of the honeycomb pattern, some of the connections break up and yield a labyrinthine pattern ($a_s = 84.0\,a_0$). The top and bottom row show 3D density distributions and 2D density cuts $n(x,y,0)$, respectively.}
	\label{fig:Pumpkin}
\end{figure*}

The honeycomb phase (Fig.~\ref{fig:PhaseDiag}(a), diamonds and squares) forms for sufficiently high atom numbers with $a_s < a_{s,c}$, where density bridges connect the central maximum and the outer ring. When another density minimum is present in the center of the trap, multiple rings with connecting density bridges and honeycomb patterns with six, seven or more density minima form. These structures feature strong density connections, facilitating superfluid flow along the honeycomb pattern \cite{Aftalion2007,Bottcher2019,Hertkorn2019,Zhang2019,Kora2019}. In combination with the crystalline structure that develops, these states form a supersolid phase \cite{Lu2015,Zhang2018}. Comparing the three-, four-, six-droplet states (stars) with the three-, four-, six-minima honeycomb states (diamonds) shown in Fig.~\ref{fig:PhaseDiag}(a) suggests that there is a symmetry between positive droplets and negative droplets on top of a background density distribution. In the infinite quasi-2D system \cite{Zhang2019}, it was shown that this is indeed a symmetry where the honeycomb structure becomes energetically favorable over the hexagonal droplet crystal beyond a critical density. We find that a similar symmetry exists in the harmonically trapped finite size system we consider here (Fig.~\ref{fig:PhaseDiag}(a), stars and diamonds). The region in which the change from droplet to honeycomb occurs is determined by an interplay between the overall density and the quantum fluctuation strength \cite{Zhang2019}.

In a window of atom numbers where the BEC-SSD boundary changes to the BEC-honeycomb boundary, the transition below $a_{s,c}$ can occur via stripes (Fig.~\ref{fig:PhaseDiag}(b), b$_2$) or honeycomb patterns deforming into stripes toward smaller $a_s$. The emergence of the stripe phase between supersolid droplets and honeycomb phases has been observed with Quantum Monte Carlo simulations \cite{Kora2019} and in a mean-field theory in a scenario where three-body interactions $\propto n^3$ \cite{Lu2015} take the stabilizing role instead of quantum fluctuations $\propto n^{5/2}$ \cite{Bulgac2002,Petrov2015,Ferrier-Barbut2016}. We have confirmed that toward larger aspect ratios, yielding larger samples (toward the thermodynamic limit), the intermediate stripe phase is enlarged in the phase diagram \cite{supmat}. When $a_s$ is further reduced, these stripes break up their connections and reenter the supersolid droplet phase (b$_1$). However, toward higher $N$ and smaller $a_s$, these stripes can curve and form overlap with neighboring stripes, representing a small region in the larger labyrinthine phase (Fig.~\ref{fig:PhaseDiag}(a) and (b), $\mathrm{b}_3$-$\mathrm{b}_6$).

This labyrinthine phase consists of elongated and curved density stripes. The amorphous spatial structure together with the strong density connections, supporting superfluid flow along the labyrinthine stripes, classify the labyrinth as a superglass. In the labyrinthine regime (Fig.~\ref{fig:PhaseDiag}(b), $\mathrm{b}_3$-$\mathrm{b}_6$) we cannot unequivocally determine the true ground state by a random initial wavefunction or by choosing a previously found low-energy state, since we find for fixed $N$ and $a_s$ many morphologically distinct labyrinthine patterns that are almost degenerate \cite{Rosensweig1983,Dickstein1993,Florence1997,Saito2009,Kawaguchi2012,Lu2015,Xi2018}, with total energy differences of a few single $\si{\hertz}$ per atom. However, we find the labyrinth states to be robust against small perturbations \cite{Dickstein1993,Florence1997,Saito2009,Xi2018}, be it in changes of scattering length or trap deformations.

With these observations about the morphologies, we now turn to the important change occurring in the phase diagram of Fig.~\ref{fig:PhaseDiag}(a), namely that the critical scattering length $a_{s,c}$ changes from rising to falling with increasing atom number. Qualitatively, the shape of the phase boundaries in Fig.~\ref{fig:PhaseDiag}(a) can be understood by noting that Eq.~\eqref{eq:EnergyFunctional} contains the three distinct scalings $\propto n$ (single-particle), $\propto n^2$ (mean-field), and $\propto n^{5/2}$ (quantum fluctuations) \cite{Petrov2015,Ferrier-Barbut2016,Bottcher2020}. While the phase diagram for low atom numbers is dominated by stabilization due to quantum pressure (kinetic energy) \cite{Wachtler2016,PitaevskiiBook2016}, the interplay between mean-field interactions and quantum fluctuations determines where $a_{s,c}$ rises quickly with atom number (Fig.~\ref{fig:PhaseDiag}(a)). For a high density, the stabilizing quantum fluctuations dominate and allow for a smaller contact repulsion with the same effective stabilization, hence the phase boundaries (including $a_{s,c}$) decrease with atom number \cite{Zhang2019,supmat}. This change of $a_{s,c}$ coincides with a peak density saturation in the ground state distributions as the honeycomb and labyrinthine phases appear for $a_s < a_{s,c}$ \cite{supmat}.  A saturating density is a defining feature of self-bound and isolated quantum droplets \cite{Petrov2015,Schmitt2016,Ferrier-Barbut2016,Chomaz2016,Baillie2016,Baillie2017,Bottcher2019droplet,Bottcher2020,Xi2020,Lee2020,Luo2020}, which develop a flat-top (spatially saturated) density distribution toward high atom numbers. The saturation signals an increasingly quantum liquid-like behavior and reduced compressibility compared to the BEC state, like for a liquid compared to a gas. Similarly for the honeycomb and labyrinthine phases, the observation of a saturating density leads to an intuitive understanding of the morphogenesis.

The effect of a saturated density in the ground state distributions for the morphogensis is best understood by following a BEC state at a high atom number through the various instability boundaries toward smaller $a_s$, as shown in Fig.~\ref{fig:Pumpkin}. Toward the atom number shown in Fig.~\ref{fig:Pumpkin}, the BEC close to $a_{s,c}$ grows and develops a shell-structure reminiscent of ultra-dense neutron stars \cite{Ravenhall1983,Pethik1995,Chamel2008,Caplan2017}. In the study of neutron stars, the occurrence of stable and nonuniform states of matter below the saturation density in the crust of the stars is known as nuclear pasta \cite{Ravenhall1983,Caplan2017}. Analogously as seen in Fig.~\ref{fig:Pumpkin}, the dense ``core" of the BEC is saturated and the density varies spatially mostly in the ``crust" of the BEC. Quantum fluctuations stabilize the core and prevent crystallization by an increasing density. Instead the system minimizes its energy by depleting density toward smaller $a_s$.  The first stage of this behavior is presented by the depleted density ring occuring in the crust of the BEC due to the inward pressure provided by the external harmonic trap (Fig.~\ref{fig:Pumpkin}, $a_s = 88.7\,a_0$). The atom number determines how close to the boundary of the BEC this depletion occurs. Toward higher $N$, the core region of the BEC grows and the depleted density ring shifts outwards. While the BEC-honeycomb transition is crossed toward smaller $a_s$ up to around $N\simeq 700 \times 10^3$ (cf. Fig.~\ref{fig:PhaseDiag}(a)), for $N \gtrsim 700 \times 10^3$ the depleted ring is located so close to the boundary (cf. Fig.~\ref{fig:Pumpkin}), that an instability similar to the fingering instability known from classical ferrofluids occurs at $a_{s,c}$ \cite{Rosensweig1983,Rosensweig1997,Dickstein1993,Jackson1994,Miranda2005,Andelman2009,Zakinyan2017}. The BEC at these high atom numbers passes through an intermediate state when $a_s$ is reduced, which we call the pumpkin state (Fig.~\ref{fig:Pumpkin}, $a_s = 88.6\,a_0$). Toward smaller $a_s$ the repulsive contact interaction and quantum fluctuations become weaker and destabilize the core region as transitions through honeycomb to labyrinthine states occur, as detailed in Fig.~\ref{fig:Pumpkin}, by a cascade of depleted density rings that form holes and wander closer to the core region.

One can connect the decrease in $a_{s,c}$ and the associated morphologies for $a_s < a_{s,c}$ to the infinite system case \cite{Zhang2019}. In the infinite system, the decrease happens roughly above a critical density at which the three phases of BEC, droplet and honeycomb are connected by a second-order phase transition \cite{NoteSecondOrderPT}. Generically below or above this critical density, the BEC is connected by a first-order transition to the honeycomb or droplet patterns in the infinite 2D system \cite{Zhang2019}. Consistent with the observations in the infinite system, here in the finite size system we find that the transition from BEC to the stripe states around the turning point of $a_{s,c}$ occurs more smoothly with no clear jump in peak density between $N\simeq 60 \times 10^3$ and $N\simeq 110 \times 10^3$ compared to the transition from BEC to the supersolid droplet or honeycomb phase at lower or higher atom numbers, respectively.

The morphogenesis of supersolid droplets for $a_s < a_{s,c}$ at low atom numbers (cf. Fig.~\ref{fig:PhaseDiag}(a)) is a special case as the system can minimize its energy by locally increasing density with the crystallization of supersolid droplets, which are not density-saturated. Studied in detail recently \cite{Schmidt2021,Hertkorn2021SSD2D}, their morphogenesis is explained by the softening of elementary excitations called angular roton modes near $a_s \simeq a_{s,c}$, which provide an angular instability and split the rotationally symmetric BEC structure into droplets.

\section{Scaling properties of quantum ferrofluids}\label{sec:ScalingProp}
The pattern formation studied above is by no means the outcome of fine-tuning of system parameters. Indeed, here we show that they are generic features of a phase diagram for dipolar quantum gases that can be discussed using dimensionless parameters and scaling laws.

We note that the ground state solution of Eq.~\eqref{eq:EnergyFunctional} is uniquely specified by the external potential parameters $\{\omega_i \}$ and the interaction parameters $(a_s,\,a_\mathrm{dd},\,N)$. In our present case, the external potential parameters correspond to the trap frequencies of the harmonic confinement, but may be left general in case of other external potentials.

We nondimensionalize Eq.~\eqref{eq:EnergyFunctional} by introducing the rescaled variables \cite{Lu2010,Bao2013,Zhang2019,Hertkorn2021SSD2D}
${\tilde{t} = t\omega_0}$, ${\tilde{\boldsymbol{r}} =  \boldsymbol{r}/x_s}$, ${\tilde{\psi} = \psi \sqrt{x_s^{3}/N}}$, with an arbitrary unit of length $x_s$ on which we base the unit of time $\omega_0^{-1} = Mx_s^2 / \hbar$ and energy $\epsilon = \hbar \omega_0$ and define the dimensionless energy functional per particle $\tilde{E} = E/N\epsilon$. After omitting the tildes the wavefunction is normalized to unity and we obtain ${E = \int \! \mathrm{d}^3r \left( \mathcal{E}_0 + \mathcal{E}_\mathrm{nl}\right)}$ with ${\mathcal{E}_0 = |\nabla \psi|^2 / 2 + V_\mathrm{ext}|\psi|^2}$, ${V_\mathrm{ext}(\boldsymbol{r}) = \sum_\alpha \gamma_\alpha^2 \alpha^2 / 2}$ for ${\alpha \in \{x,\,y,\,z\}}$, where ${\gamma_\alpha = \omega_\alpha/ \omega_0}$, and the nonlinear and nonlocal dimensionless energy density
\begin{equation}
	\mathcal{E}_\mathrm{nl}(C,D,Q) = \frac{1}{2} C |\psi|^4 +  \frac{1}{2} D |\psi|^2 (U_\mathrm{dd} * |\psi|^2) + \frac{2}{5} Q |\psi|^{5}.
\end{equation}
The dimensionless interaction strengths are given by
\begin{align}
C &= 4\pi a_s N/x_s,\label{eq:C} \\
D &= 4\pi a_\mathrm{dd} N/x_s,\label{eq:D}\\
Q &= \frac{4}{3\pi^2} \frac{C^{5/2}}{N} \left(1 + \frac{3}{2} \epsilon_\mathrm{dd}^2 \right),\label{eq:Q}
\end{align}
where $\epsilon_\mathrm{dd} = D/C$. In this formulation, the dimensionless numbers ${(C,D,Q)}$, or equivalently ${(C,D,N)}$, in addition to the external trapping parameters $\{\gamma_\alpha\}$ uniquely specify the ground state.

Since $Q$ only explicitly depends on $C$, $N$, and on the ratio $D/C$ through $\epsilon_\mathrm{dd}$, a generalization of the phase diagram (Fig.~\ref{fig:PhaseDiag}(a)) to different atomic species is straightforward. For a fixed trap geometry, we base the length unit on the dipolar length $x_s = 4\pi a_\mathrm{dd}$ and obtain ${(C,D) = (\epsilon_\mathrm{dd}^{-1}N, N)}$ \cite{Lee2020}. Therefore $Q$ is only a function of $\epsilon_\mathrm{dd}$ and $N$. Consequently in a fixed trap, the only parameters determining the type of morphology are the atom number $N$ and the relative dipolar strength $\epsilon_\mathrm{dd}$ and, for the trap discussed in Sec.~\ref{sec:Pattern}, the phase diagram generalizes to different atomic species by replacing the $a_s$-axis with $\epsilon_\mathrm{dd}^{-1}$ for any given $a_\mathrm{dd}$.

For a fixed atomic species in varying cylindrically symmetric traps, choosing $x_s = \sqrt{\hbar/M\omega_r}$ (therefore ${\omega_0 = \omega_r}$) is useful as this choice leaves only the aspect ratio $\lambda = \omega_z / \omega_0$ as an independent parameter for the external trapping potential ${V_\mathrm{ext}(\boldsymbol{r}) = (x^2 + y^2 + \lambda^2 z^2)/2}$. In this formulation, Eqs.~\eqref{eq:C}-\eqref{eq:D} reveal that the contact and dipolar interaction strengths $C \propto D \propto N \sqrt{\omega_0}$ follow the same scaling with atom number and trap frequency. Therefore $\mathcal{E}_\mathrm{nl}(C,D,0)$ is scale invariant when $N \sqrt{\omega_0}$ is kept constant \cite{Goral2000} and quantum ferrofluids in the absence of quantum fluctuations obey an important scaling property. Once a solution for a certain $(C,\,D)$ is known, an entire family of solutions with higher atom numbers and smaller trapping frequencies or vice versa has been found \cite{Goral2000,Ronen2006,Lu2010,Blakie2012}. In the presence of quantum fluctuations ($Q > 0$), the scale invariance is broken due to the explicit atom number dependence of $Q \propto C^{5/2} / N$. Therefore the strength of the stabilizing quantum fluctuations can effectively be tuned along the contours $N\sqrt{\omega_0} = \mathrm{const.}$ Such scaling properties have also proven useful for BECs interacting with an induced gravity-like interaction \cite{Odell2000, Papadopoulos2007} and one-dimensional systems \cite{Astrakharchik2018,Tylutki2020}, where they enabled the reduction of the parameter space dimension by one. In our case, the scaling behavior of $Q$ along the contours $N\sqrt{\omega_0} = \mathrm{const.}$ allows to tune the strength of the stabilization mechanism of the structured quantum ferrofluid states of matter, as we show in the following.

\begin{figure}[tb!]
	\includegraphics[trim=0 0 0 0,clip,scale=0.49]{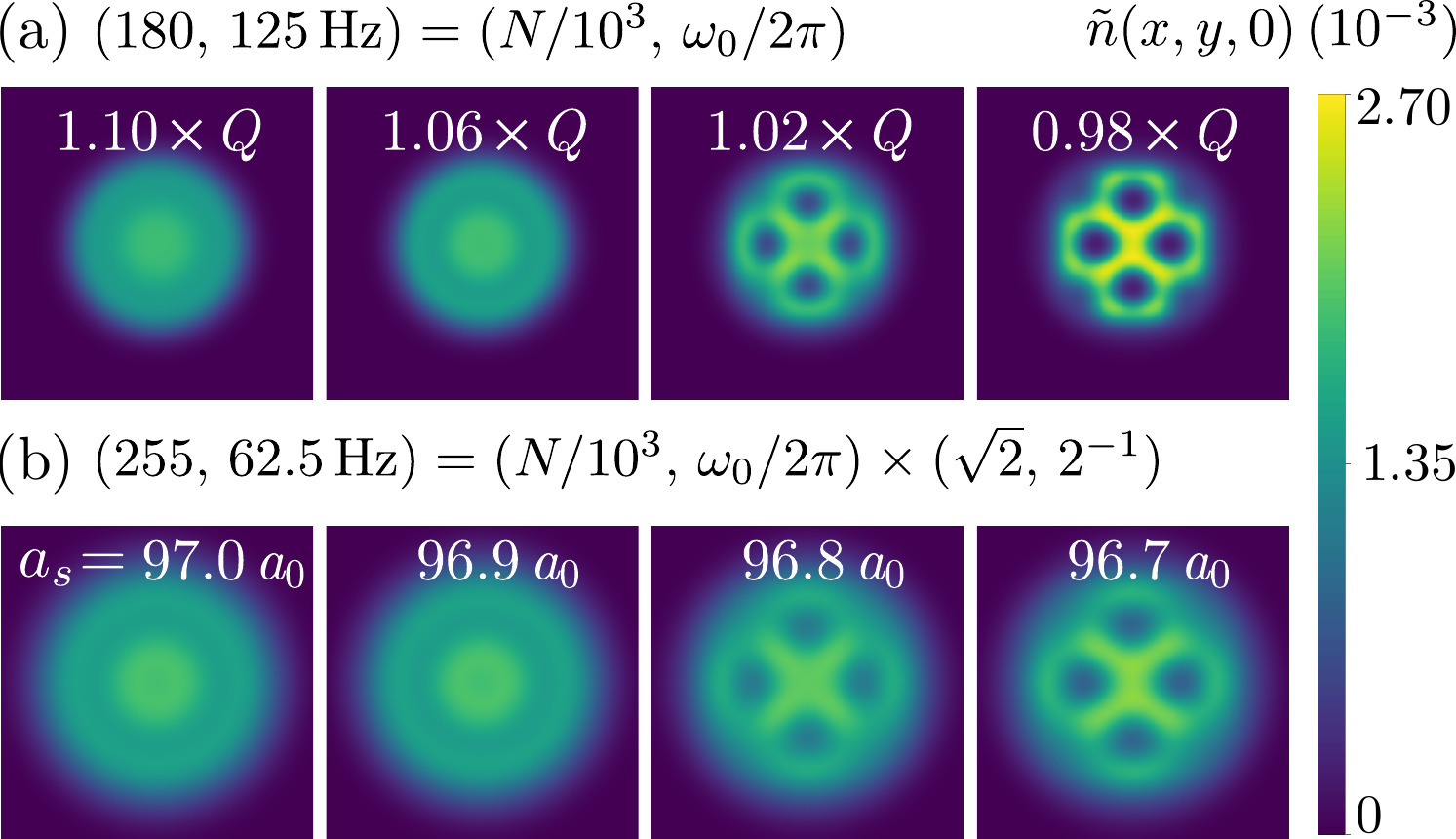}
	\caption{Tunability of quantum fluctuations. (a) Behavior of the ground state by varying $Q$, while keeping $C$ and $D$ constant. This can be realized by a scaling $s$ along contours ${N\sqrt{\omega_0} = \mathrm{const.}}$~for different atom numbers $\sqrt{s}N$ and trap geometries $\omega_0/s$. The ground state for $1.00\times Q$ was chosen with ${(N,\,\omega_0/2\pi,\,a_s) = (180 \times 10^3,\,125\,\si{\hertz},\,89.5\,a_0)}$. (b) Shows how the broken scale invariance can be used to relate similarities between different parameter regimes as an example for a trap frequency modified by a factor of two. The transition from BEC to honeycomb for a scaled atom number and correspondingly scaled trapping frequency $(\sqrt{2}N,\,\omega_0/2)$ occurs at higher scattering lengths due to the effective reduction in $Q$, which is compensated by larger $a_s$. The dimensionless density $\tilde{n} = n x_s^3/N$ at $z=0$ is shown in (a) and (b) to compare the ground state density in both trap geometries.}
	\label{fig:QFTune}
\end{figure}

In Fig.~\ref{fig:QFTune}(a), we illustrate the utility of tuning the quantum fluctuations in a quantum ferrofluid for the example of a honeycomb state. We take a four-minimum honeycomb ground state (cf. Fig.~\ref{fig:PhaseDiag}(a)) and vary the parameter $Q$ by a few percent to understand the effect of this scaling on the ground states. We see that changes in $Q$ and $a_s$ are similar since both provide a repulsive and stabilizing effect, only with a different density scaling. To this end one may note that ${C n^2 +Q n^{5/2} = \mathcal{C}(\boldsymbol{r})n^2}$ acts as an effective contact interaction, with a spatially dependent scattering length whose spatial dependence is given by $\mathcal{C}(\boldsymbol{r}) = C+Qn(\boldsymbol{r})^{1/2}$.

Figure~\ref{fig:QFTune}(b) shows how this scaling can be realized by reducing the trapping frequencies by a factor of two while keeping the aspect ratio $\lambda = 2$ constant. Due to the reduction of the stabilizing quantum fluctuations in lower confinements the BEC-honeycomb transition has shifted to higher $a_s$. Therefore at the same scattering length as in the higher confinement, the state in the lower confinement is already in the droplet regime with $a_s = 89.5\, a_0$. Toward this scattering length, the ground state in lower confinement has transitioned from the honeycomb phase through a stripe phase and finally to the supersolid droplet regime. Intuitively, $Q$ (and similarly $a_s$) controls the tendency of the density in the ground state to bond with nearby density structures. Therefore the reduced $Q$ leads to structures that bond less, the droplet state being the result of a labyrinthine state losing its tendency to bond.

Generally, similar $(C,D,Q)$ provide an efficient way to locate similar phases in the parameter space of the energy functional parametrized by the physical quantities $(a_s,a_\mathrm{dd},N)$. In particular the scaling between atom number and trap frequencies suggests that the quantum liquid states of matter shown in Figs.~\ref{fig:PhaseDiag}-\ref{fig:Pumpkin} might be observable in more tightly confined traps at experimentally accessible atom numbers \cite{Valtolina2020,Hertkorn2021SSD2D,supmat}, provided that loss mechanisms are negligible and a high optical resolution is available to resolve these fine structures. With higher trap frequencies (smaller $x_s$) one may trade off the benefit of well-separated structures (larger $x_s$) for similar ones with enhanced quantum fluctuations at smaller atom numbers. While here we only considered trap aspect ratios of two, these arguments are also valid for different cylindrically symmetric traps \cite{supmat,Bisset2016}, as we show in the following.

\begin{figure}[tb!]
	\includegraphics[trim=0 0 0 0,clip,scale=0.5]{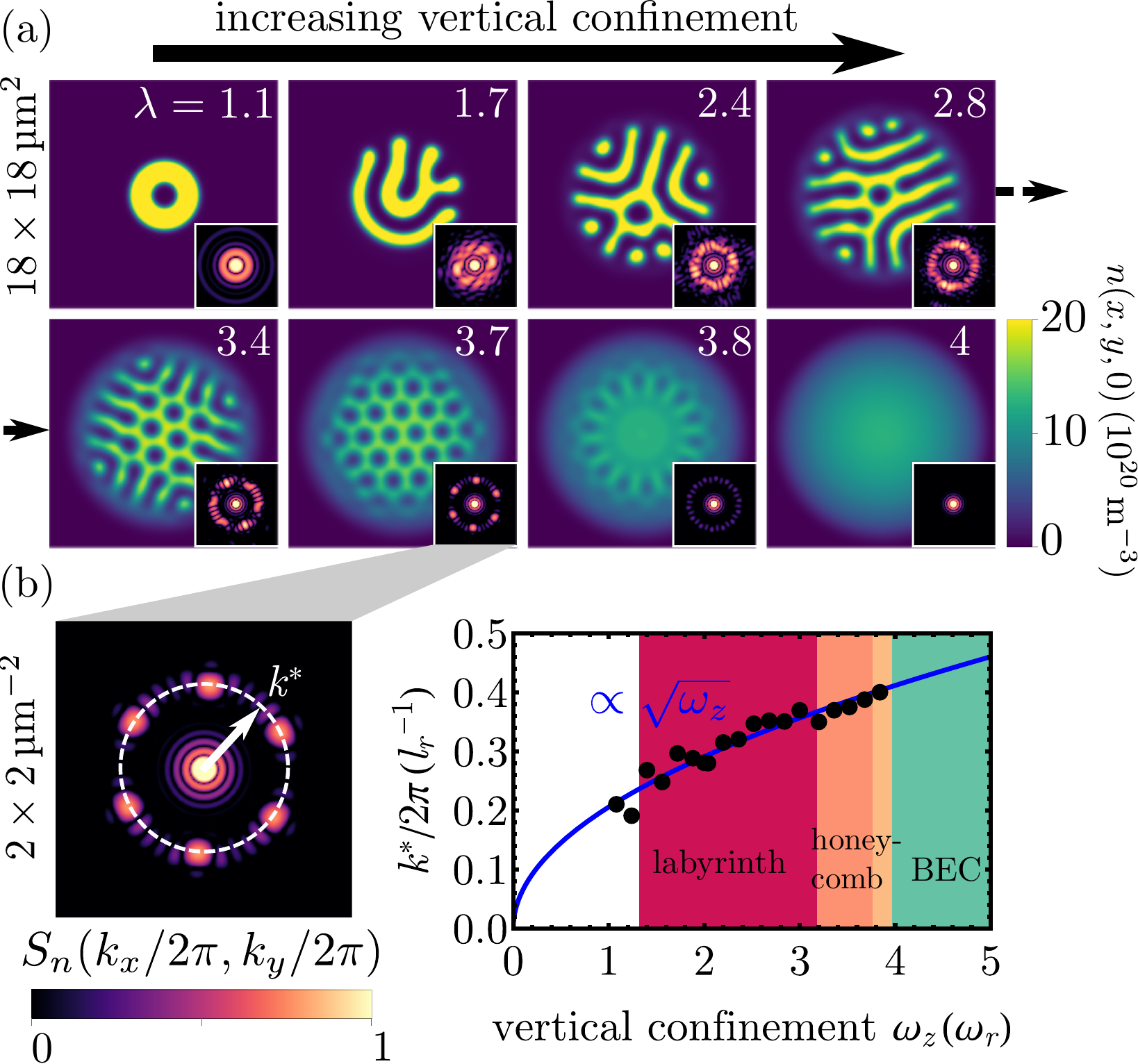}
	\caption{Vertical confinement influence on morphologies. (a) Increasing $\omega_z$ with constant ${\omega_r/2\pi = 125\,\si{\hertz}}$ yields trap geometry change induced transitions through \hbox{ring-,} \hbox{labyrinthine-,} \hbox{honeycomb-,} \hbox{pumpkin-,} and \hbox{BEC-states}. Atom number and scattering length are fixed to ${(N,\, a_s) = (500 \times 10^3,\,85\,a_0)}$. The insets in the lower right corners show the spatial power spectrum (PS) $S_n(k_x, k_y)$ in arbitrary units. The crystallinity can be seen from the diffuseness of the PS along the ring with radius ${|\boldsymbol{k}| = k^*}$. Labyrinthine states have a powdered (diffuse) PS at $k^*$, reflecting the amorphous or glassy density distribution \cite{LeBerre2002,Echeverria2020}. Toward honeycomb states, the PS concentrates in a triangular pattern indicating the increasing crystallinity. The pumpkin state ($\lambda \simeq 3.8$) PS shows more angular peaks at $k^*$ corresponding to its higher discrete rotational symmetry. (b) The characteristic momentum at radial wavevector $|\boldsymbol{k}| = k^*$ scales as ${k^* \propto 1/l_z \propto \sqrt{\omega_z}}$ and defines the characteristic spacing of the morphologies $2\pi/k^* \propto l_z$, where ${l_z = \sqrt{\hbar / M \omega_z}}$ is the harmonic oscillator length along the magnetic field direction (vertical direction). A least-squares fit to ${k^*/2\pi = c/l_z}$ as a function of vertical confinement yields ${c = 0.206(2)}$ $({l_r = \sqrt{\hbar / M \omega_r} \simeq 0.71 \, \si{\micro\meter}})$. Doubling $N$ or changing $a_s$ by $1\,a_0$ yields a similar behavior with a deviation of $c$ by less than $2\%$ (see main text).}
	\label{fig:lambdaTransition}
\end{figure}

\begin{figure}[tb!]
	\includegraphics[trim=0 0 0 0,clip,scale=0.5]{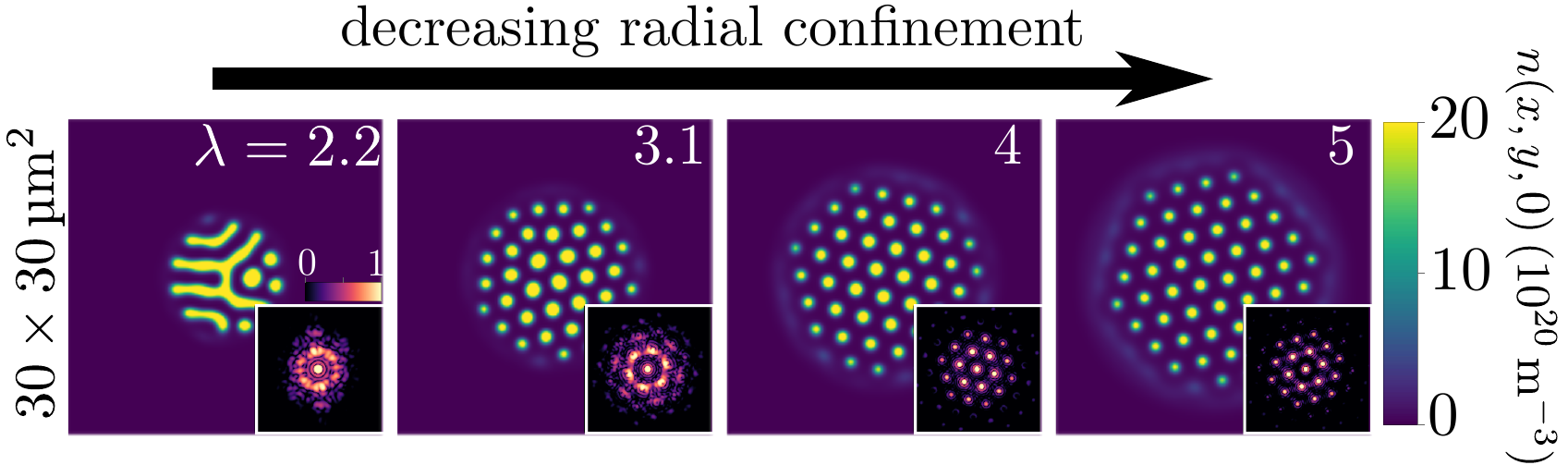}
	\caption{Radial confinement on morphologies. Reducing the radial confinement $\omega_r$ with fixed vertical confinement $\omega_z / 2\pi = 250\,\si{\hertz}$ causes the labyrinth to transition to the crystal phase. Atom number and scattering length ${(N,\, a_s) = (500 \times 10^3,\,85\,a_0)}$ are fixed as in Fig.~\ref{fig:lambdaTransition}. Across the labyrinthine to supersolid droplet transition, the characteristic momentum $k^*/2\pi \simeq 0.43\,\si{\micro\meter\tothe{-1}}$ defining the length scale of the phases stays roughly constant up to $\lambda = 5$. Toward higher $\lambda$, $k^*$ weakly increases. The most significant change is that the weight of the PS (insets) concentrates into a triangular pattern, presenting the emerging multiple Brillouin zones of the macroscopic crystalline pattern with a lattice constant $2\pi/k^*$ formed by the droplets seen in position space. As the labyrinthine patterns lose some density connections, they transition first to slightly noncylindrical droplets ($\lambda \simeq 3.1$) for which the PS still is slightly diffuse on the ring with $|\boldsymbol{k}| = k^*$ toward the pristine crystal at larger aspect ratios ($\lambda \simeq 5$).
	Reducing the radial confinement or increasing the vertical confinement (see Fig.~\ref{fig:lambdaTransition}) both increase the aspect ratio, but the effective change in the morphologies is drastically different between the two cases.}
	\label{fig:lambdaTransition_LowerRaidialConf}
\end{figure}

An interesting property of quantum ferrofluids derives from the anisotropy of the dipolar interaction, which is their geometry dependent stability 
\cite{Lahaye2008,Koch2008,Wilson2009AngularCollapse,Bisset2016,Ferrier-Barbut2018}.  The tunability of the trapping frequencies allows to investigate this geometry dependent stability continuously both theoretically and experimentally. Above, we showed that an overall scaling of trapping frequencies can be absorbed into the dimensionless interaction strengths. In the following we investigate how the morphologies are influenced by the only independent geometric parameter in the system --- the aspect ratio $\lambda = \omega_z/ \omega_r$. There is a difference between changing the aspect ratio by modifying $\omega_z$ with constant $\omega_r$ and vice versa since the magnetic field along $\hat{\boldsymbol{z}}$ breaks the symmetry between the radial and axial directions. Two cases arise, namely either a change in vertical confinement or radial confinement, as we show in Fig.~\ref{fig:lambdaTransition} and  Fig.~\ref{fig:lambdaTransition_LowerRaidialConf}, respectively.

Figure~\ref{fig:lambdaTransition}(a) shows that the ring-state in a nearly spherical trap transitions to the BEC purely by a geometric change of the trapping confinement. The state transitions through the labyrinthine phase, an increasingly macroscopically developed honeycomb phase and finally a pumpkin state. The patterns become finer as the vertical confinement increases (Fig.~\ref{fig:lambdaTransition}(a)). Analogous to the situation in classical ferrofluids confined between two plates \cite{Dickstein1993,Jackson1994,Seul1995,Florence1997}, the higher vertical confinement frustrates the morphologies more strongly and leads to their thinning. The spatial power spectrum (PS) ${S_n(k_x,k_y) = |\mathcal{F}[n(x,y,0)](k_x,k_y)|^2}$, shown in the insets of Fig.~\ref{fig:lambdaTransition}(a), reveals information about how many length scales are involved in the morphologies, the crystallinity, and the spacing (fineness) of the structures.  We have denoted ${\mathcal{F}[g](\boldsymbol{k}) = \int \! g(\boldsymbol{r})e^{i \boldsymbol{k}\cdot \boldsymbol{r}}} \mathrm{d^2}r$ as the Fourier transform of a function $g$. Since the states have no modulation along $z$ the PS of the cut suffices to analyze the structures. The PS is concentrated radially around a single characteristic momentum $|\boldsymbol{k}| = k^*$. This single radial concentration shows that there is only a single characteristic length scale in the morphologies, corresponding to $2\pi/k^*$. The spacing (fineness) of the structures can be seen in the absolute value of $k^*$ as a function of vertical confinement. Figure~\ref{fig:lambdaTransition}(b) reveals that the spacing scales as $2\pi/k^* \propto l_z$, where ${l_z = \sqrt{\hbar / M \omega_z}}$ is the harmonic oscillator length along the magnetic field direction.

This scaling behavior is known from the roton momentum $k_\mathrm{rot}$, defining the characteristic momentum at which the dispersion relation of a dipolar BEC shows a distinct roton minimum \cite{Santos2003,Jona-Lasinio2013,Baillie2015}. The collective excitations associated to this minimum, the roton modes, are precursors to a structural phase transition when the roton minimum softens near zero excitation energy. Representing the dominant fluctuations driving this transition \cite{Hertkorn2021,Schmidt2021,Hertkorn2021SSD2D}, the roton modes carry their length scale, the roton wavelength $\lambda_\mathrm{rot} = 2\pi/k_\mathrm{rot}$, over into the newly emerging ground state structure and provide its characteristic structural length scale $2\pi/k^*$. The fact that the characteristic length scale across a structural phase transition can be interpreted to originate from softening or energetically low-lying excitations on the higher-symmetry-side of the transition is a generic result of linear stability analysis in nonlinearly interacting systems, such as classical ferrofluids \cite{Dickstein1993,Rosensweig1983} or nonlinear optics \cite{Baio2020,Labeyrie2014,Zhang2021,Zhang2018,Maucher2016} and is therefore general beyond the situation in quantum ferrofluids \cite{Archer2008,Heinonen2019}. In the supersolid droplet regime, this behavior has been thoroughly studied recently \cite{Roccuzzo2019,Hertkorn2019,Natale2019,Hertkorn2021,Hertkorn2021SSD2D}. Figure~\ref{fig:lambdaTransition} shows that this scaling behavior persists from the  BEC state to the honeycomb phase, throughout the multistable labyrinthine phase to the ring state.

Relating the domain spacing to the roton momentum suggests that the coefficient $c = 0.206(2)$ for the characteristic momentum $k^*/2\pi = c/l_z$ mostly depends on chemical potential and maximum density in the system \cite{Santos2003,Jona-Lasinio2013,Baillie2015,Chomaz2018}. As the density is saturated for the labyrinthine and honeycomb phases, the chemical potential varies weakly with atom number in these regimes. Therefore, $c$ varies weakly with atom number and yields a robust characterization of the fineness of the structures for a given interaction strength and trap geometry. We have repeated the analysis shown in Fig.~\ref{fig:lambdaTransition} with a different scattering length $a_s = 84\,a_0$ and atom numbers ${N=\{700,\ 1000\} \times 10^3}$ and find that $c$ varies by less than than $ 2\% $ at these different parameters. Toward lower atom numbers, the peak density and chemical potential become more sensitive to interaction parameters and trapping frequencies and $c$ is generally a function of these parameters. However, the scaling $k^* \propto 1/l_z$ remains.

Figure~\ref{fig:lambdaTransition_LowerRaidialConf} shows the behavior of the morphologies with decreasing radial confinement with a fixed vertical confinement $\omega_z / 2\pi = 250\,\si{\hertz}$ for the same $a_s$ and $N$ as in Fig.~\ref{fig:lambdaTransition}. Instead of a transition from labyrinthine phase through honeycomb and pumpkin states toward the BEC (Fig.~\ref{fig:lambdaTransition}), one finds in Fig.~\ref{fig:lambdaTransition_LowerRaidialConf} that the labyrinthine phase loses its density connections and transitions into the crystalline droplet phase. The PS (insets in Fig.~\ref{fig:lambdaTransition_LowerRaidialConf}) shows that the characteristic momentum $k^*$ does not change during the transition. These observations can be understood as follows.

Equations~\eqref{eq:C}-\eqref{eq:Q} with $x_s = \sqrt{\hbar/M\omega_r}$ show that a decreasing radial confinement leads to a reduction in the dimensionless interaction strengths similar to a decreasing atom number \cite{supmat}. In the phase diagram (Fig.~\ref{fig:PhaseDiag}) this decrease corresponds to a crossing of the labyrinthine-SSD boundary at constant $a_s$, explaining the labyrinthine to supersolid droplet transition seen in Fig.~\ref{fig:lambdaTransition_LowerRaidialConf}. Since decreasing $\omega_r$ additionally leads to an increase of the natural length scale $x_s \propto 1/\sqrt{\omega_r}$ and aspect ratio $\sqrt{\lambda} \propto x_s$, the transition is not exactly equivalent to a change in atom number but corresponds to a trajectory through four-dimensional parameter space $(C,D,Q,\lambda)$ \cite{supmat}. As the spacing of the structures at constant $(C,D,Q)$ decreases as ${2\pi/ k^* \propto 1/\sqrt{\lambda}}$ (Fig.~\ref{fig:lambdaTransition}), but for the case of decreasing $\omega_r$ the natural length scale $x_s \propto \sqrt{\lambda}$ expands at the same rate, these two effects roughly balance and lead to a constant $k^*$.

Finally, we note that a change in the aspect ratio combined with a change in atom number according to $\lambda \rightarrow \infty,\, N\rightarrow \infty,\, n_0 = \mathrm{const.}$ corresponds to systems approaching the thermodynamic limit \cite{Santos2003,Jona-Lasinio2013,Baillie2015,Cinti2014,Macri2013,Macri2014,Saccani2012}. Accordingly one expects quantum ferrofluids to form more macroscopic structures toward larger aspect ratios. Repeating the calculation for the phase diagram toward larger aspect ratios, we indeed find that the structures become more macroscopic and that the morphologies discussed in Sec.~\ref{sec:Pattern} prevail \cite{supmat}.

\section{Conclusion and outlook}\label{sec:Conclusion}
In conclusion we identify new quantum liquid forms of matter in quantum ferrofluids beyond the supersolid droplet regime. We have shown a general phase diagram of quantum ferrofluids in an oblate trap, which features supersolid droplets at low densities and labyrinthine, honeycomb, and pumpkin states toward higher densities. The emergence of these morphologies can be traced back to the increasingly dominant role of quantum fluctuations toward higher densities, providing the underlying stabilizing mechanism. The strength of this stabilization can be tuned by adjusting the overall trapping confinement. Due to the anisotropy of the dipolar interaction, the morphologies can be transformed into one another by a simple adjustment of the trap aspect ratio. Squeezing the quantum ferrofluid morphologies along the magnetic field direction reveals that the characteristic length scale of the morphologies follows the same scaling behavior as the roton wavelength known from ordinary BEC states.

The labyrinthine states hint at a large degeneracy of the ground state within the framework of an effective mean-field description. This calls for a more elaborate theory beyond the effective description in this labyrinth phase, which however is beyond the scope of the current work. In particular, an interesting possibility is that the various labyrinthine morphologies we find to be degenerate in our effective description might actually be selected upon by quantum fluctuations \cite{Sachdev1992}.

Another direction worth investigating is to obtain further insight into the dominant collective excitations giving rise to the honeycomb and labyrinthine morphologies. A linear stability analysis similar to studies on the BEC to supersolid droplet transition \cite{Roccuzzo2019,Natale2019,Hertkorn2019,Hertkorn2021SSD2D} may allow identification of modes characteristic of the supersolid or superglass nature of these patterns.

We anticipate that an extension of our study to molecules \cite{Carr2009,Valtolina2020} with tunable electric dipole moments could reveal further interesting phases in regimes where strong correlations and the granular nature of matter play an important role \cite{Liu2008,Archer2008,Cinti2010}.

\textit{Note added.} Upon submission of the present work, we became aware of a related and very recent preprint \cite{Zhang2021phases}.

\section*{Acknowledgments}
We thank Hans Peter B\"uchler, Detlef Lohse, J\"orn Dunkel and Vili Heinonen for inspiring discussions. M.G. and M.Z. acknowledge funding from the Alexander von Humboldt Foundation. T.L. acknowledges funding from the European Research Council (ERC) under the European Union’s Horizon 2020 research and innovation programme (Grant agreement No. 949431). This work is supported by the German Research Foundation (DFG) within FOR2247 under Pf381/16-1 and Bu2247/1, Pf381/20-1, FUGG INST41/1056-1 and the QUANT:ERA collaborative project MAQS.

\bibliography{refs} 

\begin{thebibliography}{156}
\bibitem [{Not()}]{NoteSecondOrderPT}%
  \BibitemOpen
  \bibinfo {note} {{The authors considered an expansion of the eGPE energy
  functional up to fourth order in a density wave amplitude. Generally, it
  cannot a priori be ruled out that the transition could still be of first
  order if higher orders of the amplitude were considered.}}\BibitemShut
  {Stop}%
\end{thebibliography}

\clearpage

\setcounter{section}{0}
\setcounter{figure}{0}
\renewcommand{\figurename}{Supplementary Figure}
\renewcommand{\thefigure}{S\arabic{figure}} 
\renewcommand{\theequation}{S.\arabic{equation}}
\newcounter{SFfig}
\renewcommand{\theSFfig}{S\arabic{SFfig}}

\section*{Supplementary material}
\subsection{Phase diagram}
We find ground states using conjugate gradient techniques \cite{Modugno2003,Ronen2006,Antoine2017,Antoine2018}. When searching the ground state from a random initial wavefunction for the phase diagram shown in Fig.~\ref{fig:PhaseDiag}, we use gradient noise \cite{Perlin1985}. We typically find a faster convergence with gradient noise compared to white noise or gaussian states. The mean-field dipolar potential is effectively calculated using Fourier transforms, where we use a spherical cutoff for the dipolar potential. The cutoff radius is set to the size of the simulation space such that there is no spurious interaction between periodic images \cite{Goral2002,Ronen2006,Lu2010}.

\begin{figure}[tb!]
	\includegraphics[trim=0 0 0 0,clip,scale=0.4]{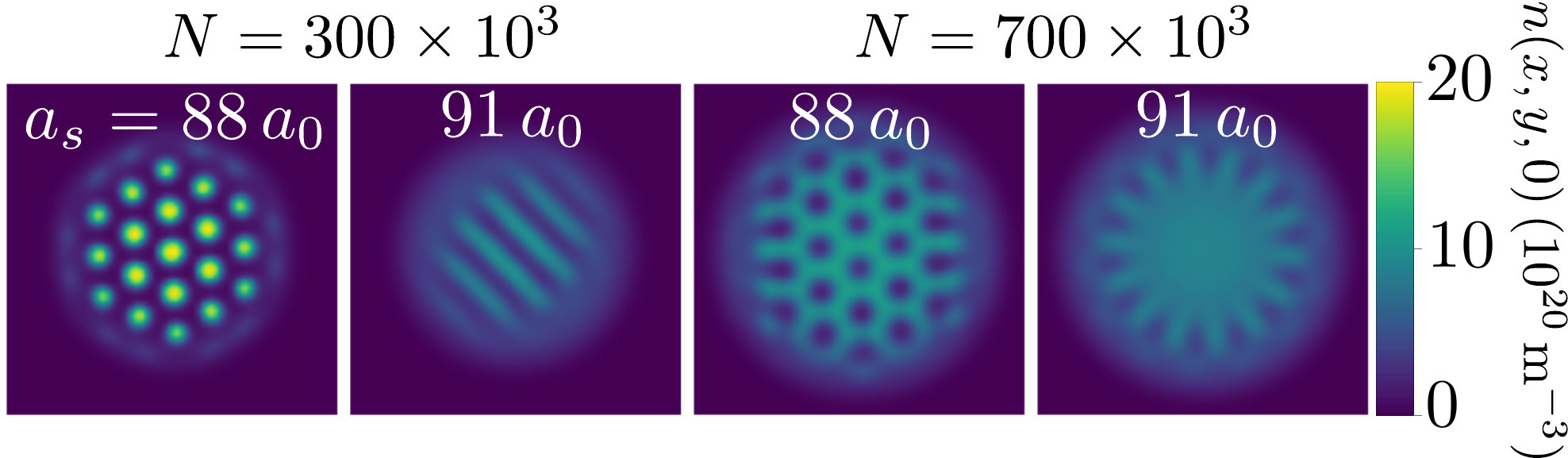}
	\caption{Morphologies in a trap with trapping frequencies ${\omega/2\pi= (83.3,\,83.3,\,250)\,\si{\hertz}}$ ($\lambda$ = 3). Shown are 2D density cuts $n(x,y,0)$ in a field of view of $30 \times 30 \, \si{\micro\meter\tothe{2}}$. The phases do not require a specific trap geometry and can be found at different interaction strengths and atom numbers for different trap geometries. Macrosopic structures form toward higher aspect ratios.}
	\label{fig:lambda3}
\end{figure}

In order to understand the behavior of the morphologies towards the thermodynamic limit, we recalculate the phase diagram (see Fig.~\ref{fig:PhaseDiag}(a)) for an aspect ratio of $\lambda = 3$ by keeping $\omega_z/2\pi = 250\,\si{\hertz}$ constant and reducing the radial trapping frequencies to $\omega_r/2\pi = 83.3\,\si{\hertz}$. The droplet, labyrinth, honeycomb and pumpkin phases can be found in the new phase diagram as well and the relative location of the phase boundaries are similar to Fig.~\ref{fig:PhaseDiag}(a). Examples of the droplet, stripe, honeycomb and pumpkin states for an aspect ratio of $\lambda = 3$ are shown in Fig.~\ref{fig:lambda3}. Compared to smaller aspect ratios, the morphologies have expanded and become more macroscopic, as expected. More droplets, stripes, honeycomb minima and fingers of the pumpkin state form. The boundary described by $a_{s,c}$ has shifted to higher atom numbers and scattering lengths, and the rate is smaller with which $a_{s,c}$ decreases toward higher atom numbers above the critical atom number $N \simeq 200 \times 10^3$. The shift of the boundaries in lower radial confinements can be intuitively understood by considering that quantum fluctuations are reduced (see Sec.~\ref{sec:ScalingProp}), and therefore higher scattering lengths and atom numbers are required to reach similar patterns. The fact that in different trap geometries the overall structure of the phase diagram is similar, in particular that the superglass and supersolid states of matter prevail, shows that no fine-tuning of the trap geometry or atom numbers is necessary to observe these structures. The morphologies are not fine-tuned states but rather phases of matter in the complex phase diagram of quantum ferrofluids. Furthermore, the scaling relation provided in Sec.~\ref{sec:ScalingProp} provide an intuitive understanding of the changes induced on the boundaries between these phases by an overall change in trapping frequencies or the atom number.

\begin{figure}[tb!]
	\includegraphics[trim=0 0 0 0,clip,scale=0.525]{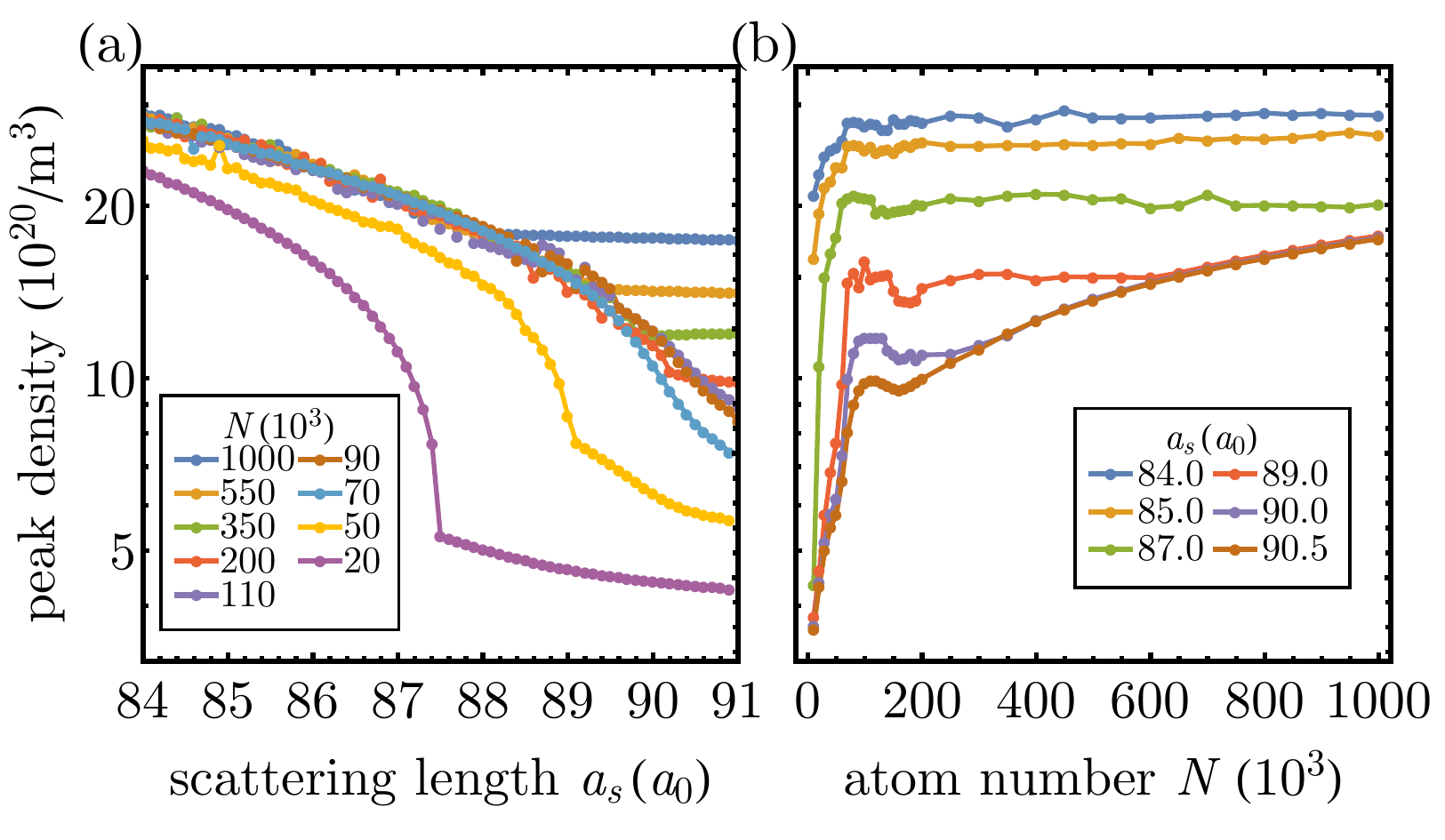}
	\caption{Peak density in the phase diagram shown in Fig.~\ref{fig:PhaseDiag} as a function of scattering length at fixed atom numbers (a) and as a function of atom number for fixed scattering lengths (b).}
	\label{fig:figSup1}
\end{figure}

For completeness of the discussion regarding the peak density in the main text, we show the peak density $n_0$ along vertical and horizontal cuts of the phase diagram in the main text (Fig.~\ref{fig:PhaseDiag}(a)) in Fig.~\ref{fig:figSup1}. Figure~\ref{fig:figSup1}(a) shows that the peak density has a jump at small atom numbers, when the supersolid droplet regime is entered and that the discontinuity becomes smaller toward higher atom numbers. Above ${N \simeq 120 \times 10^3}$, where the honeycomb phase separates the BEC phase from the labyrinth, the critical scattering length $a_{s,c}$ decreases with increasing atom number. In the labyrinthine phase, there are fluctuations in peak density as the scatttering length is varied since labyrinths with different forms can have slightly different peak densities. However, toward smaller scattering lengths they follow the same general functional form $n_0(a_s)$ regardless of the atom number, which indicates that the peak density is saturated in the labyrinthine phase. The saturation can also be seen in Fig.~\ref{fig:figSup1}(b) as a function of atom number for fixed scattering lengths. In the BEC regime (${a_s = \{90,\, 90.5\}\,a_0}$) the peak density rises relatively quickly up to ${N\simeq 120 \times 10^3}$, where the behavior of the critical scattering length $a_{s,c}$ changes, as described in the main text. Slightly below this atom number, the stripe phase appears as an intermediate region between the honeycomb and the supersolid droplet (SSD) phase (see Fig.~\ref{fig:PhaseDiag}). Above ${N\simeq 120 \times 10^3}$ in the BEC regime, the peak density grows significantly slower compared to the initial increase and only weakly depends on the scattering length (cf. Fig.~\ref{fig:figSup1}(a)). Spatially, the core region of the BEC close to the honeycomb transition is roughly density saturated (see Fig.~\ref{fig:Pumpkin}) and grows slowly when the atom number is increased as shown in Fig.~\ref{fig:figSup1}(b). One can see from  the peak density with $a_s = 89\,a_0$, where the honeycomb phase is entered and exited as a function of atom number, that in the honeycomb phase the peak density is saturated and when it is exited, follows the same behavior of the BEC. At smaller scattering lengths, one can see that the peak density still grows in the droplet regime when the atom number is increased, but is saturated in the stripe and labyrinthine phases.

A change of the peak density behavior can be seen in the BEC phase close to the instability boundary where the atom number is high enough to support a density maximum in the center of the trap, surrounded by a ring of depleted density near the boundary (Fig.~\ref{fig:PhaseDiag}(a), circles). Increasing the atom number from there on mainly leads to an overall growth of the BEC structure while maintaining the depleted density ring near its boundary. The overall peak density in the BEC still grows for higher atom numbers, but at a smaller rate \cite{supmat}. 

Towards smaller scattering lengths in the honeycomb and labyrinthine phases, the depleted density needs to redistribute itself among the remaining density connections, leading to a moderate increase of density in these structures. This process leads to the transition from honeycomb to labyrinthine states, as some density connections weaken sufficiently toward low scattering lengths to break up. The states in the honeycomb and labyrinthine phases do not increase their peak density toward higher atom numbers but only grow in size and change their morphology. Along the density lines in the honeycomb and labyrinthine phases, the density remains spatially almost flat.

Honeycomb and labyrinthine phases with their macroscopically saturated density distribution realize a quantum liquid that is even further extended in space than the previously studied isolated and self-bound quantum droplets \cite{Bulgac2002,Petrov2015,Schmitt2016,Kadau2016,Ferrier-Barbut2016,Chomaz2016,Baillie2016,Baillie2017,FerrierBarbut2018GetThin,FerrierBarbut2019,Bottcher2019droplet,Bottcher2020,Xi2020,Lee2020,Luo2020}. Despite their saturated density $n_0$, these phases are still ultradilute compared to strongly interacting or ordinary liquids \cite{FerrierBarbut2018GetThin,FerrierBarbut2019} as the gas parameter in the entire phase diagram of Fig.~\ref{fig:PhaseDiag}(a) stays below $n_0 a_s^3 \lesssim 3 \times 10^{-4}$.

There is an analogy to the situation of elongated supersolids \cite{supmat}, where a decrease in critical scattering length beyond a critical density has also been noticed when quantum fluctuations are included in the description \cite{Chomaz2019,Blakie2020variational,Blakie2020supersolelongate}. In elongated geometries, the transition from BEC to droplets is smooth in an intermediate density regime \cite{Blakie2020supersolelongate} and the decrease in critical scattering length is only observed when the LHY correction is included \cite{Blakie2020variational}. Beyond the intermediate atom number regime, it is seen that the density modulation forms first around the outer boundary and moves inwards for decreasing $a_s$ \cite{Chomaz2019}. Analogously in the oblate trap, we find that the density modulation at low atom numbers occurs in the center of the trap (blood cell and subsequent droplet formation) and at higher atom numbers the density minimum of the blood cell moves outwards and bridges form between the central maximum and the outer ring, yielding the honeycomb phase, as shown in Fig.~\ref{fig:PhaseDiag}. The smooth transition in an intermediate atom number regime can also be seen in the round trap as we discussed in the main text, as well as the infinite system \cite{Zhang2019}.

We point out a similarity to self-assembling collodial systems with competing interactions \cite{Liu2008,Archer2008}. In these systems, the phase diagram has a similar basic structure as shown in Fig.~\ref{fig:PhaseDiag} \cite{Archer2008}, and minimum-energy configurations can show the formation of shell structures \cite{Liu2008}, in which multiple depleted density regions form in a honeycomb pattern.

We conclude with final remarks regarding the phase diagram on the transition between the honeycomb and labyrinthine phases.

When setting a previously found ground state across the honeycomb-labyrinthine transition, we find that the discrete rotational symmetry of the honeycomb state persists to smaller scattering lengths compared to the state one finds when searching from a random initial state. In some cases, honeycomb structures can lose their outer connections toward smaller scattering length which yields droplets that surround circularly symmetric density rings. When comparing these states with stripe and labyrinth states that we obtain by searching for the ground state from a random initial wavefunction, we find that the ring states are typically a few $\si{\hertz}$ up to ten $\si{\hertz}$ higher in total energy per atom, which indicates that these are metastable states originating from a hysteresis. Nonetheless we find that these states may be relevant for future work, as we performed real-time simulations where we slowly ramp the scattering length across the transition and find that these metastable ring states can be long-lived (we have evolved these states up to $120\, \si{\milli\second}$ after the ramp is complete and find them to be stable) and might therefore be experimentally observable.

\subsection{Reduced units and scaling properties}\label{sec:RedUnits}
As described in the main text, we define the dimensionless variables ${\tilde{t} = t\omega_0}$, ${\tilde{\boldsymbol{r}} =  \boldsymbol{r}/x_s}$, ${\tilde{\psi} = \psi \sqrt{x_s^{3}/N}}$ to nondimensionalize the energy functional  \cite{Lu2010,Blakie2012,Bao2013,Zhang2019,Lee2020,Hertkorn2021SSD2D}. Here, $\omega_0^{-1}$ and $x_s$ can at first be taken as arbitrary quantities with units of time and length, respectively. For Schr\"odinger-like equations it is convenient to define the energy and time units as $\epsilon = \hbar^2 / M x_s^2$ and $\omega_0^{-1} = M x_s^2 / \hbar$  based on the unit of length $x_s$.

A significant consequence is that the contact and dipolar interaction terms (Eqs.~\eqref{eq:C}-\eqref{eq:D}) for a given atomic species are only ever modified by the product $N\sqrt{\omega_0}$. This is a result of the contact and dipolar interaction both being quadratic in $\psi$ and since the three-dimensional convolution ($\propto \mathrm{d^3}r$) of the dipolar interaction ($U_\mathrm{dd} \propto 1/r^3$) stays invariant when scaling space by a factor of $x_s$. With other nonlocal interactions following different power law behaviors, for instance in systems with an induced gravity-like interaction ($\propto (1/r) * |\psi|^2$), different scaling properties with the atom number can be exploited \cite{Odell2000,Papadopoulos2007}. In the main text we focused on the behavior $C\propto D \propto N\sqrt{\omega_0}$ in order to describe the change of the ground state when different atom numbers and trapping frequencies are considered. However, the reduced units are also generally useful to discuss the behavior of different atomic or molecular species for fixed trapping frequencies by considering the full dependence $C \propto a_sN/x_s$, $D \propto a_\mathrm{dd}N/x_s$. These expressions include the scattering and dipolar lengths, as well as the mass $M$ that influences $x_s \propto 1/\sqrt{M}$ and $a_\mathrm{dd} \propto \mu_m^2 M$.

Let us consider systems with higher masses or stronger magnetic dipole moments compared to $^{162}\mathrm{Dy}$, yielding larger $D \propto \mu_m^2 M^{3/2} N$. Scaling for $s<1$ the atom number and mass $N \to sN$, $M \to M/s^2$ increases $D \to D/s^2$, $Q \to Q/s$ and leaves $C$ unchanged. To obtain the same $\epsilon_\mathrm{dd}$, higher scattering lengths are required, which leads to enhanced quantum fluctuations. Similar to the scaling ${N\sqrt{\omega_0} = \mathrm{const.}}$~we discussed previously, the quantum fluctuations $Q$ are enhanced along the contours ${\mu_m^2 M^{3/2} N=\mathrm{const.}}$ for larger $\mu_m$ or $M$ at smaller $N$.

Another use of the dimensionless interaction strengths is to obtain an intuitive understanding of the geometry dependence of dipolar BECs discussed in Figs.~\ref{fig:lambdaTransition}-\ref{fig:lambdaTransition_LowerRaidialConf} of the main text. The fact that the labyrinthine-SSD transition for smaller $\omega_r$ at constant $\omega_z$ occurs (cf. Fig.~\ref{fig:lambdaTransition_LowerRaidialConf} in the main text), can be understood by considering $(C,D,Q)$ as a coordinate system on which geometric transformations are performed as $\omega_r$ or $N$ is changed. While Fig.~\ref{fig:lambdaTransition} shows the behavior of the system in the full parameter space $(C,D,Q,\lambda(\omega_z))$ along the fourth and independent dimension $\lambda = \omega_z/ \omega_r$, Fig.~\ref{fig:lambdaTransition_LowerRaidialConf} shows behavior along a trajectory ${\omega_r \mapsto \mathcal{T}(\omega_r) =  (C(\omega_r),D(\omega_r),Q(\omega_r),\lambda(\omega_r))}$ through four-dimensional parameter space.

As for the scaling properties discussed, it is useful to consider the change on the coordinate system $(C,D,Q)$ induced by the transformation $\omega_r \rightarrow \omega_r / s$ and one obtains $(C,D,Q) \rightarrow (C,D,Q/s^{3/4})/\sqrt{s}$. Interpreting this change as a geometric transformation \cite{Han1997,Dirac1949}, it is a contraction combined with a squeeze mapping that squeezes the quantum fluctuations. For $s>1$, the overall interaction strengths decrease, the quantum fluctuation are additionally reduced (``squeezed" closer to zero), and the natural length scale increases $x_s \rightarrow \sqrt{s} x_s$. We compare to the transformation $N \rightarrow N / \sqrt{s}$, yielding ${(C,D,Q) \rightarrow (C,D,Q/s^{1/4})/\sqrt{s}}$ and see that it is the same contraction combined with a weaker squeeze mapping.

From this comparison, we see that a decrease in atom number corresponds to a similar change in the interaction parameters $(C,D,Q)$ compared to a change in the radial trapping confinement, except that in addition $\lambda$ changes. This observation provides an intuitive understanding of the state changing from labyrinthine to droplets. During this transformation $\mathcal{T}(\omega_r)$ is a path similar to moving toward smaller atom numbers at constant $\epsilon_\mathrm{dd}$ in the phase diagram of Fig.~\ref{fig:PhaseDiag}(a), exiting the labyrinthine phase and entering the droplet phase. Since the natural length scale increases $x_s \rightarrow \sqrt{s} x_s$, one can consider the change in radial confinement as effectively moving toward smaller atom numbers and simultaneously evaluating the ground state on rescaled spatial coordinates, that increase as $\propto \sqrt{s}$. Were all interaction parameters $(C,D,Q)$ kept constant during the change of $\lambda$, the situation shown in Fig.~\ref{fig:lambdaTransition}(a) would occur and the characteristic length scale of the morphologies would change as $\propto \lambda \propto 1/\sqrt{s}$. Therefore, these two effects roughly balance and one observes a transition from labyrinth to droplet state, where the droplets keep their lattice spacing roughly constant and merely grow radially outwards toward a macroscopic droplet crystal.
\end{document}